\documentclass[aps,prd,twocolumn,showpacs,floatfix]{revtex4}
\newcommand{\be}{\begin{equation}}
\newcommand{\ee}{\end{equation}}

\newcommand{\mnras}{Mon. Not. Roy. Astr. Soc. }

\newcommand{\aj}{Astron. J.}
\begin{document}

\title{Best median values for cosmological parameters}

\author{Matts Roos}
\affiliation{Department of Physical Sciences, Division of High
Energy Physics, University of Helsinki, Helsinki}

\date{\today}

\begin{abstract}
Our procedure to obtain best values for cosmological parameters from
five recent multiparameter fits is as follows. We first study the
values quoted for $r,\ \alpha_s,\ w+1 (w_0+1),\ w_1$ and $\Omega_k$,
arriving at the conclusion that they do not differ significantly
from zero, and their correlations to other parameters are
insignificant. In what follows they can be therefore ignored. The
neutrino mass sum $\Sigma m_{\nu}$ also does not differ
significantly from zero, but since neutrinos are massive their sum
must be included as a free parameter. We then compare the values
obtained in five large flat-space determinations of the parameters
$\Sigma m_{\nu},\ \omega_b,\ \omega_m,\ h,\ \tau,\ n_s,\ A_s$ and
$\sigma_8$. For these we compute the medians and the 17-percentile
and 83-percentile errors by a described procedure.

\end{abstract}
 \maketitle

\section{Introduction}
Since July 2004 five large analyses
\cite{Seljak2,Eisenstein,MacTavish,Sanchez,Sievers} of cosmological
parameters have appeared (we call them papers P1, P2, P3, P4 and P5)
based on partly overlapping data. They represent a wealth of new
information which, however, may appear contradictory to the outsider
and difficult to digest. Since each analysis determines parameter
values separately for different subsets of data and for several
choices of parameter spaces, the results are {\em legio}. For
instance, these five papers quote altogether 56 different values for
the density of matter in the Universe, $\omega_m$. The parameter
values exhibit notable differences due to differing choices of
priors, methods of analysis, simulation variance and statistical
fluctuations. The purpose of this review is to compare these five
analyses, and to try to extract as reliable values as possible.
Although the authors of the analyses may be unlikely to believe in
the results of anyone else, theoreticians do need recommended
values.

\section{Data sets}
All analyses quote results from fits to different subsets and
combinations of data sets. This is of course of importance in order
to test different sets for consistency or for correlations. However,
since the purpose of the present analysis is to obtain as accurate
information about parameter values as possible, we consider only
fits to maximal data sets.

All the five papers P1--P5 use the CMB $\langle TT\rangle$
\cite{Hinshaw} and $\langle TE\rangle$ \cite{Kogut} power spectra on
large angular scales from the first year Wilkinson Microwave
Anisotropy Probe (WMAP) observations. P1 \cite{Seljak2} combines
this with the constraints from the Sloan Digtal Sky Survey (SDSS)
galaxy clustering analysis \cite{Tegmark}, the SDSS galaxy bias
analysis at $z=0.1$ \cite{Seljak1}, and the SNIa constraints
\cite{Riess}. A distinctive feature of the P1 analysis is the
inclusion of Ly-$\alpha$ forest data \cite{McDonald}. P1 mentions
explicitly that inclusion of power spectra from the CMB observations
on small angular scales by the Cosmic Background Imager (CBI) in
2004 \cite{Readhead}, the Very Small Array (VSA) \cite{Dickinson},
and the Arc\-minute Cosmology Bolometer Array Receiver (ACBAR)
\cite{Kuo} do not affect their results much, nor does the inclusion
of the 2 degree Field Galaxy Redshift Survey (2dFGRS) \cite{Cole}
galaxy clustering power spectrum.

P2 \cite{Eisenstein} uses an SDSS "Main sample" \cite{Strauss} of
galaxy clustering data which is of earlier date than the sample used
in P1. The distinctive feature of the P2 analysis is the inclusion
of spectroscopic data for a recent sample of 46 748 luminous red
galaxies (LRG) in the range $0.16<z<0.47$ which show a significant
acoustic peak in the redshift-space two-point correlation function
(related to the galaxy power spectrum). P2--P5 do not include
Ly-$\alpha$ forest data, some of them mentioning that the
information is still susceptible to systematic errors.

P3 \cite{MacTavish} uses in addition to WMAP \cite{Hinshaw,Kogut}
the following sets of CMB data: the BOOMERANG $\langle TT\rangle$
power spectra, the BOOMERANG $\langle TE\rangle  + \langle EE\rangle
+ \langle BB\rangle$ polarization spectra, and the $\langle
TT\rangle$ power spectra of DASI \cite{Halverson}, VSA
\cite{Dickinson}, ACBAR \cite{Kuo}, MAXIMA \cite{Hanany}, and CBI
2004 data \cite{Readhead}. The shapes of the SDSS and 2dFGRS power
spectra \cite{Tegmark,Cole} are used as constraints, but not the
amplitudes. In addition the SNIa "gold" set \cite{Riess} is
included.

P4 \cite{Sanchez} relies on either the 2dFGRS \cite{Cole} or the
SDSS \cite{Tegmark} power spectrum of galaxy clustering , the WMAP
data \cite{Hinshaw,Kogut}, but in contrast to P1 they also use the
$\langle TT\rangle$ power spectra of VSA \cite{Dickinson}, ACBAR
\cite{Kuo}, and CBI 2004 data \cite{Readhead}. We use the parameter
values derived from 2dFGRS data rather than SDSS, because P4
represents 2dFGRS.

P5 \cite{Sievers} analyzes exclusively CMB data: $\langle TT\rangle$
from \cite{Hinshaw,Readhead,Halverson,MacTavish} and in particular
all the $\langle TE\rangle  + \langle EE\rangle$ polarization data
now available \cite{MacTavish,Kogut,Halverson,Sievers}. They do not
use LSS constraints.

It would have been interesting to include also the recent analysis
of cosmic shear \cite{Kilbinger}. However this would not have been
consistent because the parameter space analyzed is quite different
from that of papers P1 -- P5.

\section{Parameter spaces}
The most general parameter space explored in the five papers is
13-dimensional \be \textbf{p}=(\omega_b, \omega_m, \Omega_k, \Sigma
m_{\nu}, \tau, h, \sigma_8, b, w, w_1, n_s, A_s, \alpha_s, r)\ . \ee
The normalized matter density parameter $\Omega_m$ does not appear
explicitly in this list because it is given by $\Omega_m=\omega_m/
h^2$ where $h$ is the Hubble parameter in units of 100 km s$^{-1}$
Mpc$^{-1}$. Similarly, $\Omega_b=\omega_b/ h^2$ is the normalized
baryon density parameter. One has $\omega_m =\omega_{dm}+\omega_b$
and hence $\Omega_m =\Omega_{dm}+\Omega_b$, where $\Omega_{dm}
=\omega_{dm}/h^2=\Omega_{cdm}+\Omega_{\nu}$ is the density parameter
of dark matter, sometimes denoted $\Omega_c$ for CDM. The density
parameter of dark energy is then $\Omega_{\Lambda} =1- \Omega_m-
\Omega_k$, where $\Omega_k$ is the curvature or vacuum density. The
ratio of pressure to energy density for dark energy is
$w=w_0+w_1(1-a)$, where $a$ is the scale parameter of the Universe.
$\sigma_8$ is the rms linear mass perturbation in $8h^{-1}$ Mpc
spheres.

The fraction of dark matter that amounts to neutrinos is $f_{\nu} =
\Omega_{\nu}/\Omega_{dm}$. Assuming that the neutrinos are Majorana
particles with standard freeze-out, the sum of the three neutrino
masses is $\Sigma m_{\nu}=94.4\ \omega_{dm}f_{\nu}$ eV.

The parameters $\tau,\ n_s,\ \alpha_s,\ A_s,\ r,\ b$ describe
fluctuation properties: the scalar spectral index is $n_s$, the
scalar amplitude is $A_s$, the tensor spectral index $n_t$ is not
included because there is no information, the ratio of tensor to
scalar amplitudes is $r=A_t/A_s$, and the running of $n_s$ with $k$
is $\alpha_s$. The parameter $\tau$ measures the Thomson scattering
optical depth to decoupling, and $b$ is the bias factor describing
the difference in amplitude between the galaxy power spectrum and
that of the underlying dark matter. For $b$ no value is quoted.

\section{Method of analysis}
Since the data sets are to a large extent overlapping they are not
independent.  Moreover, parts of the reported errors are systematic
and not stochastic, so different data sets cannot be summarized
using statistical methods, either frequentist or Bayesian. The
breakdown into data subsets differing in methods of data sampling,
statistical analysis, and simulation demonstrate notable variations
in the parameter values (see for example Chu et al. \cite{Chu}). We
clearly want to use results from as large data compilations as
possible, not from subsets.

Many parameters are determined with quite low precision, the
published $1\sigma$ errors are large, the correlations likewise
insignificant, and the dependences on the choices of data subsets,
priors, and parameter spaces are statistically weak. The tendency in
the cosmological literature is to draw optimistic conclusions from
$1\sigma$ and $2\sigma$ confidence regions obtained by
marginalization in many-parameter spaces. We think that parameter
estimates from the five data sets, at their present state of
accuracy, can well be combined without sophisticated methods. We
then proceed as follows.

Each measurement $i$ of a parameter $a$ we represent by a Gaussian
ideogram   with central value $a_i$, symmetrized error $\delta a_i$,
and unit area [in least squares fitting one weighs the area by
$(\delta a_i)^{-2}$]. We sum these ideograms, and define our best
values as medians (as do for instance P1 and P5), not means. In lieu
of variances we define our errors as the 17-percentiles and the
83-percentiles of the sum ideogram. The most notable effect is then
that these "errors" are always larger than the smallest individual
error in the data set, whereas in statistical averaging errors
always come out smaller than the smallest error entering. One may
take this as an estimate of systematic errors for which there anyway
does not exist any statistical prescription.
\begin{table}
\caption{\label{tab:table1}Comparison of empirically defined errors}
\begin{ruledtabular}
\begin{tabular}{cccc}
$a_1\pm\delta a_1$ & $a_2\pm\delta a_2$ & Our method & $|a_2 - a_1|/2$ \\
 & & & $+\delta a_1$\\
$1\pm 5.0$ & $2\pm 5.0$ & $1.5\pm 4.80$ & $1.5\pm 5.5$ \\
$1\pm 1.0$ & $2\pm 1.0$ & $1.5\pm 1.08$ & $1.5\pm 1.5$ \\
$1\pm 0.5$ & $2\pm 0.5$ & $1.5\pm 0.71$ & $1.5\pm 1.0$ \\
$1\pm 0.2$ & $2\pm 0.2$ & $1.5\pm 0.58$ & $1.5\pm 0.7$ \\
$1\pm 0.5$ & $2\pm 1.0$ & $1.33^{\ +1.09}_{\ -0.67}$  \\
\end{tabular}
\end{ruledtabular}
\end{table}

Some numerical examples in Table I demonstrate the properties of
this method. In comparison we give the results for another empirical
method in which $|a_2 - a_1|/2$ is taken as an estimate of a
systematic error which is added linearly to the statistical error
$\delta a_1=\delta a_2$. (Another prescription is needed for
asymmetric errors.) The last line in the Table exemplifies the
combination of data with unequal errors by our method. The
penultimate line exemplifies data which clearly are in
contradiction. No such flagrant contradictions appear in the data
that we shall combine.

Of course no probability measure or statement of statistical
confidence can be associated with our type of errors. Whenever we
mention $1\sigma$ errors or confidence limits (CL), they refer to
quantities in the data sets.

\section{Parameter values}
\subsection{$r,\ \alpha_s,\ w,\ w_0,\ w_1,\ \Omega_k$}
Let us first discuss parameters for which the
theoretical motivation is model dependent and the evidence
insignificant.

P1 quotes three upper limits (95\% CL) for the parameter $r$, the
smallest value being $r<0.36$ in a 7-parameter fit with
$\alpha_s=0,\ \Sigma m_{\nu}=0$ and $w=-1$. The effect of including
$r$ as a seventh free parameter is that the information on the basic
six parameters is degraded -- their errors increase. Their central
values change only within the quoted 1$\sigma$ errors, thus $r$ is
not significantly correlated to the basic parameters. The largest
correlation is with $n_s$, still an increase of only about 1$\sigma$
(this is discussed in detail in P1 \cite{Seljak2}). When any of the
parameters $\alpha_s,\ \Sigma m_{\nu},\ w$ is allowed to vary, the
parameter space becomes 8- or 9-dimensional resulting in a further
deterioration of the information on the basic parameters as well as
on $r$.

P3 quotes $r<0.36$ and P4 quotes $r<0.41$ from fits in the same
7-parameter space. The inclusion of $r$ as a free parameter degrades
slightly the information on some basic parameters. As in the case of
P1, $r$ is not significantly correlated to any other parameter
except slightly to $n_s$, at the level of about 1$\sigma$.

P2 and P5 do not quote any value for $r$. Thus there is no
significant evidence for $r\neq 0$, only for a commonly acceptable
limit $r<0.4$.

The parameter $\alpha_s$ is determined by P1 and P3. P1 quotes fits
in two different parameter spaces:

\noindent $r<0.45,\ \alpha_s=-0.006^{+0.012}_{-0.011},\ w=-1$, and

\noindent $r<0.45,\ \alpha_s=-0.011\pm 0.012,\
w=-0.91^{+0.08}_{-0.09}$.

\noindent In neither case is $\alpha_s$ significantly different from
zero.

P3 quotes $\alpha_s=-0.051^{+0.027}_{-0.026}$ when $w=-1$ in rather
marked disagreement with P1. Because of this conflict we choose not
to combine it with P1. The effect of including $\alpha_s$ in the P1
and P3 fits increases the errors of the basic parameters
considerably. In P1 no significant correlations are found, in P3
there are some correlations at the 1$\sigma$ level. That the running
of $n_s$ is poorly determined is not surprising since we shall see
later that also $n_s$ is poorly determined.

To proceed with $w$ we take $r= 0$ and $\alpha_s=0$. With this
assumption four values of $w$ have been determined:

\noindent P1: $-0.99\pm 0.09\ \ $, P2: $-0.80\pm 0.18\ \ $, P3:
$-0.94^{+0.09}_{-0.10}$, P4: $-0.85^{+0.18}_{-0.17}$.

The inclusion of $w$ as a seventh parameter in P1 and P3 has very
little effect on the errors and central values of the basic six,
more so in P4, but the correlations are still insignificant. One
notices in P1, however, that keeping $r$ and $\alpha_s$ free
improves the $w$ errors and reduces its central value by slightly
more than 1$\sigma$.

Including $w$ as a fifth parameter in P2, degrades the errors of
their basic four parameters, and changes their central values by
less than 1$\sigma$.

The above values can then be used to define the median and errors
\be w=-0.93^{+0.13}_{-0.10}\ . \ee

P1 also determines the parameters $w_0,\ w_1$ in the combination
$w=w_0+w_1(1-a)$, with the result

\noindent $w_0=-0.98\pm 0.19,\ w_1=0.05^{+0.83}_{-0.65}$.

\noindent Thus there is no significant information indicating $w\neq
-1$ or $w_1\neq 0$ at present. In a Bayesian study of the need for
either of the parameters $w$ or $n_s$ (using a modified P1 data set)
Mukherjee et al. \cite{Mukherjee} conclude that theories with more
parameters are not disfavored, but that the improvement does not
warrant the additional complexity in the theory. Thus we choose
results from 7-parameter fits which assume $w=-1$. {\em A fortiori}
$w_1$ is useless.

The available determinations of the curvature para\-meter $\Omega_k$ for the
case $w=-1$ are

\noindent P2: $-0.010\pm 0.009$, P3: $-0.027\pm 0.016$, P4:
$-0.074^{+0.049}_{-0.052}$.

The general tendency is confirmed that the inclusion of $\Omega_k$
degrades the information on the basic parameters, also the
correlations are insignificant with one exception: there is a
positive correlation between $\Omega_k$ and $h$, most visible in P3.
The above set of data define the median and errors \be
\Omega_k=1-\Omega_0=-0.023^{+0.017}_{-0.050}.\ee

We conclude that there is very little information on curvature at
present, and that $h$ to some extent takes its r\^{o}le.

In the sequel we shall only use parameter values determined under
the assumptions of flat space and $r=\alpha_s=w+1=0$. This maximizes
the information on the basic parameters, and it does not entirely
neglect the effects of possibly non-vanishing $\Omega_k$ or $r$
because of the presence of the slight ($\Omega_k,\ h$) and ($r,\
n_s$) correlations.

\subsection{Neutrinos}

Although there are only upper limits determined for $\Sigma m_{\nu}$
there is clear evidence from neutrino oscillations that the
neutrinos have mass. Therefore it is more motivated to include
$f_{\nu}$ or $\Sigma m_{\nu}$ in the fits than any of the parameters
discussed in the previous subsection. There is neutrino information
in all papers except P5 and P2; the latter notes that $n_s$ has the
same effect as massive neutrinos. P3 and P4 obtain relatively small
values for $n_s$ and higher neutrino masses, but this is not a very
significant correlation.

The Ly-$\alpha$ forest data present more difficulties than the other
data sets in the form of nuisance parameters which have to be
marginalized over and systematic errors which have to be estimated.
The Ly-$\alpha$ forest data are orthogonal to the other data in the
sense that they are responsible for much of the improvement on the
primordial spectrum shape and amplitude:  $\sigma_8$  is more
accurately determined and the neutrino mass limit is much tighter.
P1 admits that more work is needed, reflecting the sceptical
attitude in the other papers.

The 95\% C.L. results for  $\Sigma m_{\nu}$ in units of eV are

\noindent P1: $<0.42\ (<1.54$ without Ly-$\alpha$), P3: $<1.07$, P4:
$<1.16$.

\noindent Our recommendation is cautiously \be \Sigma m_{\nu}<1.1\
\rm{eV}\ . \ee

In the following we shall use those parameter values in P1, P3 and
P4 that were obtained with $\Sigma m_{\nu}$ as the seventh free
parameter. The P2 data, obtained with $\Sigma m_{\nu}=0$, shall be
used in a special way to be described shortly. One notes in the P1,
P3 and P4 fits that some of the parameters are quite insensitive to
whether $\Sigma m_{\nu}$ is free or not: this is true for
$\omega_b,\ \omega_m,\ n_s$ and $A_s$. For these parameters we
consider the values obtained in P5 to be sufficiently unbiased to be
used here.

\subsection{$\omega_b,\ \Omega_b,\ \omega_m,\ \Omega_m,\ h$}
Let us compare P1 \cite{Seljak2} and P2 \cite{Eisenstein}. The SDSS
"Main sample" in P2 is not exactly the same as what P1 calls "all".
The effects of the spectroscopic LRG data on the Main sample are
quoted, generally they are small. For flat $w=-1$ cosmology the LRG
data change the parameters obtained from the Main sample by the
amounts

\noindent $\Delta \omega_m=-0.004,\ \Delta\Omega_m=-0.007,\ \Delta
h=-0.004,\ \Delta n_s=-0.017$.

\noindent Assuming that the LRG data would cause the same
corrections to the P1 results, we concoct a set of LRG-corrected
results that we call P1+LRG.

The $\omega_b$ values found in the five analyses are very consistent, uncorrelated with other parameters and robust against different choices of parameter sets and priors, as pointed out by P4 \cite{Sanchez}. They are

\noindent P1: $2.36\pm 0.09$, P3: $2.24^{+0.08}_{-0.09}$, P4:
$2.24^{+0.12}_{-0.11}$, P5: $2.31\pm 0.13$.

We can combine these with the BBN value \cite{BBN} $2.2\pm 0.2$, to obtain
values of the median and errors \be
10^2\omega_b=2.28^{+0.12}_{-0.13}\ . \ee

The parameter $\omega_m$ is determined better than
$\Omega_m=\omega_m/h^2$, thus the route to $\Omega_m$ is to use
$\omega_m$ and $h$. The values for $\omega_m$ have often been
obtained in this way. Also, since WMAP constrains $\omega_m$ rather
than $\Omega_m$, one then avoids a large ($\Omega_m,\ h^2$)
correlation. We turn first to the Hubble parameter $h$ which is
measured to be

\noindent P1+LRG: $0.706^{+0.023}_{-0.022}$, P3:
$0.648^{+0.039}_{-0.038}$, P4: $0.691\pm 0.038$.

P3 and P4 exhibit clearly the effect of making the neutrino mass
variable, a change of $-3.1\%$ and $-6.4\%$ respectively. This is a
measure of neutrino mass bias in P2 and P5 that we therefore do not
include. The chosen input yields the median and errors \be
h=0.687^{+0.034}_{-0.047}\ , \ee \noindent in flat space, in
remarkably good agreement with our constrained fit in 1998
\cite{Nevalainen} $h=0.68\pm 0.05$, based on Cepheid distances only.

From the above values of $\omega_b$ and $h$ we derive the fraction
of baryonic energy in the Universe \be
\Omega_b=0.048^{+0.005}_{-0.004}\ . \ee

Turning now to the $m$ and $dm$ parameters, the available input data
are the P1+LRG value $\Omega_m=0.277\pm 0.025$, the P2 value
$\omega_m \equiv \omega_{dm}+ \omega_b=0.142\pm 0.005$, and the
$\omega_{dm}$ values

\noindent  P3: $0.126\pm 0.007$, P4: $0.110\pm 0.006$, P5: $0.112\pm
0.011$.

Combining these values with $h$ and $\omega_b$ in the most efficient
way, we find the median and errors \be \omega_m=0.139\pm 0.011,\ \ \
\ \ \Omega_m=0.286^{+0.030}_{-0.028} . \ee

\noindent This is in excellent agreement with what WMAP quoted in
2003, $\Omega_m=0.27\pm 0.04$ \cite{Hinshaw}. It follows from this
that the dark energy content in flat space is \be
\Omega_{\Lambda}=0.714^{+0.028}_{-0.030}\ , \ee  and the fraction of
dark matter \be \Omega_{dm}=0.238^{+0.030}_{-0.028}\ . \ee

\subsection{$\tau,\ n_s,\ A_s,\ \sigma_8$}
The parameters $\tau$ and $\Omega_k$ are significantly correlated
\cite{Sanchez}, so if $\Omega_k$ is taken to be zero, also $\tau$
will be small. P1 and P2 use the prior $\tau<0.3$ which appears a bit
tight; even so P1 obtains the highest value for $\tau$. Since all
papers use WMAP data, they inherit the problems with $\tau$
appearing different in the Northern and Southern hemispheres
\cite{Wibig,Larson}. Thus this parameter is neither reliably nor
precisely determined. The values that can be quoted are

\noindent P1: $0.185^{+0.052}_{-0.046}$, P3:
$0.108^{+0.049}_{-0.047}$, P4: $0.143^{+0.076}_{-0.071}$, P5:
$0.147\pm 0.085$.
\texttt{twocolumn}
\begin{table*}
\caption{\label{tab:table2}Recommended values of parameters}
\begin{ruledtabular}
\begin{tabular}{cccc}
Parameter & Definition & Median value & 83-/17-percentile, \\
& & & or 95\% CL limit \\
$10^2\omega_b$ & $10^2\times$ baryon density & 2.28 \footnotemark[1]
& $+0.12 / -0.13$
\\
$\Omega_b=\omega_b/h^2$ & Normalized baryon density & 0.048
\footnotemark[3]& $+0.005 / -0.004$
\\
$\omega_m$ & Total matter density & 0.139 \footnotemark[1] & $\pm 0.011$ \\
$\Omega_m=\omega_m/h^2$ & Normalized matter density & 0.286 \footnotemark[3]& $+0.030/ -0.028$ \\
$\Omega_{dm}=\Omega_m-\Omega_b$ & Normalized dark matter density & 0.238 \footnotemark[3]& $+0.030/ -0.028$ \\
$\Omega_{\Lambda}=1-\Omega_m-\Omega_k$ & Normalized dark energy density & 0.714 \footnotemark[3]& $+0.028/ -0.030$ \\
$h$ & Hubble parameter [100 km/s Mpc]& 0.687 \footnotemark[1] & $+0.034/ -0.047$ \\
$\Sigma m_{\nu}$ & Neutrino mass sum & & $<1.1$ eV \\
$\tau$ & Thomson scattering optical depth to decoupling & 0.147 \footnotemark[1] & $+0.068/ -0.064$ \\
$n_s$ & Scalar spectral index & 0.962 \footnotemark[1] & $+0.030/ -0.027$ \\
$\ln(10^{10}A_s)$ & Scalar fluctuation amplitude & 3.12 \footnotemark[1] & $+0.14/ -0.12$ \\
$\sigma_8$ & RMS linear mass perturbation in $8h^{-1}$ Mpc spheres & 0.81 \footnotemark[3]& $+0.09/ -0.14$ \\
$r=A_t/A_s$  & Ratio of tensor to scalar amplitude fluctuations & & $ <0.4$ \footnotemark[2] \\
$\alpha_s=\rm{d}n_s/\rm{d}k$ & Running scalar index & -0.011 \footnotemark[2]& $\pm 0.012$ \\
$w$ & Dark energy EOS & -0.92 \footnotemark[2]& $+0.17/ -0.12$ \\
$\Omega_k$ & Normalized vacuum density & -0.023 \footnotemark[2]& $+0.017 / -0.050$ \\
\end{tabular}
\end{ruledtabular}
\footnotetext[1]{$\ $From 7-parameter fits including $\Sigma
m_{\nu}$.} \footnotetext[2]{$\ $From alternative 7-parameter fits
which exclude $\Sigma m_{\nu}$.} \footnotetext[3]{$\ $Alternative or
derived parameter.}
\end{table*}
From this, the values of the median and errors are \be
\tau=0.147^{+0.068}_{-0.064}\ . \ee

The parameter $n_s$ is measured by P1, P3, P4 and P5. The LRG data in P2
do not yield independent information on $n_s$ and $\omega_m$, so P2
adds them in the form of a constraint
$\omega_m=0.130(n_s/0.98)^{-1.2}\pm 0.011 $ that we do not use. The available data are then

\noindent P1+LRG: $0.972^{+0.026}_{-0.023}$, P3: $0.95\pm 0.02$, P4:
$0.957^{+0.031}_{-0.029}$, P5: $0.975\pm 0.038$.

From this, the values of the median and errors are \be
n_s=0.962^{+0.030}_{-0.027}\ , \ee \noindent thus $n_s$ is
marginally less than 1.0.

There are three measurements of the amplitude $A_s$, quoted in
the form $\ln(10^{10}A_s)$:

\noindent P3: $3.1\pm 0.1$, P4: $3.11^{+0.15}_{-0.14}$, P5: $3.17\pm 0.16$.

From this, the values of the median and errors are \be
\ln(10^{10}A_s)=3.12^{+0.14}_{-0.12}\ . \ee

The final parameter in this survey is $\sigma_8$. P2 has fixed its
value at 0.85, the other experiments quote values exhibiting a rather large variance,

\noindent P1: $0.890^{+0.035}_{-0.033}$, P3: $0.74\pm 0.08$, P4:
$0.678^{+0.073}_{-0.072}$, P5: $0.849\pm 0.062$.

From this, the values of the median and errors are \be
\sigma_8=0.81^{+0.09}_{-0.14}\ . \ee

Since $\sigma_8$ is rather strongly correlated with $\tau$ and
$\Omega_m$, a more accurate determination of $\Omega_m$ could have
been obtained if all experiments had agreed to fix $\sigma_8$ and
$\tau$ at some common values.

\section{Conclusions}
By combining results from five large data analyses
\cite{Seljak2,Eisenstein,MacTavish,Sanchez} we find that no
significant values can be obtained at present for $r,\ \alpha_s,\
w,\ w_0,\ w_1$, and that the values for $\Omega_k$ and $n_s$ are
only marginally significant, consistent with flat space and no tilt.
This agrees well with a Bayesian study \cite{Mukherjee} (using much
of the P1 data) which concludes that $w$ and $n_s$ do not improve
any fits.

We consider it important to include $\Sigma m_{\nu}$ among the
varied parameters since neutrino oscillation results have proven
that neutrinos have mass. Yet the fits we quote yield only an upper
limit for $\Sigma m_{\nu}$.

We have shown that it is possible to recommend best values for
$\omega_b,\ \omega_m,\ h,\ \tau,\ n_s,\ A_s,\ \sigma_8,\ \Omega_k$
and $n_s$, rather regardless of whether $\tau$ and $\sigma_8$ are
taken to be fixed or not; fixing would improve the accuracy of
several parameters. Our results are collected in Table II.

\end{document}